\documentclass[usenatbib]{mn2e}
\usepackage{graphicx}
\usepackage{ifthen}
\usepackage{amsmath}
\usepackage{url}
\usepackage{rotating}
\usepackage{xcolor}

\def\ltsima{$\; \buildrel < \over \sim \;$}
\def\lta{\lower.5ex\hbox{\ltsima}}
\def\gtsima{$\; \buildrel > \over \sim \;$}
\def\simgt{\lower.5ex\hbox{\gtsima}}
\def\kms{{\rm\,km \; s^{-1}}}

\def\AA{$\; \buildrel \circ \over {\rm A}$}

\def\s{\ifmmode \widetilde \else \~\fi}
\def\={\overline}

\def\spose#1{\hbox to 0pt{#1\hss}}

\def\lta{\mathrel{\spose{\lower 3pt\hbox{$\mathchar"218$}}
     \raise 2.0pt\hbox{$\mathchar"13C$}}}
\def\gta{\mathrel{\spose{\lower 3pt\hbox{$\mathchar"218$}}
     \raise 2.0pt\hbox{$\mathchar"13E$}}}
\def\Dt{\spose{\raise 1.5ex\hbox{\hskip3pt$\mathchar"201$}}}    % upper case
\def\dt{\spose{\raise 1.0ex\hbox{\hskip2pt$\mathchar"201$}}}    % lower case

\def\dotsfill{\leaders\hbox to 1em{\hss.\hss}\hfill}

\loadboldmathitalic 
\title[The Pristine Dwarf-Galaxy survey III]{The Pristine Dwarf-Galaxy survey -- III. Revealing the nature of the Milky Way globular cluster Sagittarius II}
\author[N. Longeard et al.] {Nicolas Longeard$^{1}$, Nicolas Martin$^{2,3}$, Rodrigo A. Ibata$^{2}$, Else Starkenburg$^{4}$, 
\newauthor Pascale Jablonka$^{1,5}$, David S. Aguado$^{6}$, Raymond G. Carlberg$^{7}$, Patrick C\^ot\'e$^{8}$, 
\newauthor Jonay I. Gonz\'alez Hern\'andez$^{9,10}$, Romain Lucchesi$^{1}$, Khyati Malhan$^{11}$, Julio F. Navarro$^{12}$, 
\newauthor Rub\'en S\'anchez-Janssen$^{13}$, Guillaume F. Thomas$^{8}$, Kim Venn$^{12}$, Alan W. McConnachie$^{8}$ \\
$^{1}$ Laboratoire d'astrophysique, \'Ecole Polytechnique F\'ed\'erale de Lausanne (EPFL), Observatoire, 1290 Versoix, Switzerland\\
$^{2}$ Universit\'e de Strasbourg, CNRS, Observatoire astronomique de Strasbourg, UMR 7550, F-67000 Strasbourg, France\\
$^{3}$ Max-Planck-Institut f\"ur Astronomy, K\"onigstuhl 17, D-69117, Heidelberg, Germany\\
$^{4}$ Leibniz Institute for Astrophysics Potsdam (AIP), An der Sternwarte 16, 14482 Potsdam, Germany\\
$^{5}$ GEPI, Observatoire de Paris, Universit\'e PSL, CNRS, Place Jules Janssen, F-92195 Meudon, France\\
$^{6}$ Institute of Astronomy, University of Cambridge, Madingley Road, Cambridge CB3 0HA, UK \\
$^{7}$ Department of Astronomy \& Astrophysics, University of Toronto, Toronto, ON M5S 3H4, Canada \\
$^{8}$ NRC Herzberg Astronomy and Astrophysics, 5071 West Saanich Road, Victoria, BC V9E 2E7, Canada\\
$^{9}$ Instituto de Astrof\'isica de Canarias, V\'ia L\'actea, 38205 La Laguna, Tenerife, Spain \\
$^{10}$ Universidad de La Laguna, Departamento de Astrof\'isica, 38206 La Laguna, Tenerife, Spain \\
$^{11}$ The Oskar Klein Centre for Cosmoparticle Physics, De- partment of Physics, Stockholm University, AlbaNova, 10691 Stockholm, Sweden \\
$^{12}$ University of Victoria, 3800 Finnerty Rd, Victoria, BC, V8P 5C2, Canada \\
$^{13}$ STFC UK Astronomy Technology Centre, Royal Observatory, Blackford Hill, Edinburgh, EH9 3HJ, UK \\
}

\date{\today}
\begin{document} 
\maketitle 
\begin{abstract}
We present a new spectroscopic study of the faint Milky Way satellite Sagittarius~II. Using multi-object spectroscopy from the Fibre Large Array Multi Element Spectrograph, we supplement the dataset of \citet{longeard20} with 47 newly observed stars, 19 of which are identified as members of the satellite. These additional member stars are used to put tighter constraints on the dynamics and the metallicity properties of the system. We find a low velocity dispersion of $\sigma_\mathrm{v}^\mathrm{SgrII} = 1.7 \pm 0.5$ km s$^{-1}$, in agreement with the dispersion of Milky Way globular clusters of similar luminosity. We confirm the very metal-poor nature of the satellite ([Fe/H]$_\mathrm{spectro}^\mathrm{SgrII} = -2.23 \pm 0.07$) and find that the metallicity dispersion of Sgr~II is not resolved, reaching only $0.20$ at the $95$\% confidence limit. No star with a metallicity below $-2.5$ is confidently detected. Therefore, despite the unusually large size of the system (r$_h = 35.5 ^{+1.4}_{-1.2}$ pc), we conclude that Sgr~II is an old and metal-poor globular cluster of the Milky Way.
\end{abstract}
 

\section{Introduction}
The faint and metal-poor satellites of massive galaxies are among the oldest structures of the Universe, whether they are dwarf galaxies or globular clusters (\citealt{white_rees_78}, \citealt{beasley02}, \citealt{mo10}). Because the faintest, low-mass galaxies are, in a $\Lambda$CDM context, dominated by dark matter (DM), they can be the key to constraining the nature of the DM particle and more broadly the properties of our Universe (\citealt{klypin99}, \citealt{geringer-sameth15a}, \citealt{bullock17}, \citealt{nadler19}). The old age of the dwarf galaxies further implies that they have hosted Population III stars (\citealt{ishiyama16}), while their low mass and metallicity suggest that the pollution from successive supernovae is limited in these systems (\citealt{dekel86}, \citealt{frebel_norris15}). As a consequence, they are unique laboratories to study and quantify the different evolutionary pathways of stellar formation (\citealt{roederer16}, \citealt{webster16}, \citealt{ji19}). 

On the other hand, globular clusters are usually not thought to be DM-dominated \citep{moore96} and the faintest of them are often modelled as simpler systems with a single stellar population (\citealt{gratton07}, \citealt{carretta09}, \citealt{cohen10}, \citealt{willman_strader12}). They are important for various reasons. They give us insight in the mode of stellar formation in the early Universe and its similarities and differences with star formation today (\citealt{gieles18},\citealt{gratton19}). The close link of their properties to those of their host galaxies has become clear over the years: for example, the number of clusters are related to the properties of their host galaxies (\citealt{brodie_huchra91}, \citealt{blakeslee97}). They are thereby direct probes of their galaxy' properties and therefore indirect probes for cosmology (\citealt{cote02_cluster}, \citealt{villaume19}, \citealt{riley20}). Old globular clusters are also witnesses of the build-up of massive galaxies halos, which are thought to have formed by disrupting and accreting faint stellar systems (\citealt{beasley18}, \citealt{pfeffer18}). They can also offer a unique insight into the observational validation of stellar population models \citep{chantereau16}.

The diversity of the properties of the MW faint satellites has been progressively unveiled over the past decades. For dwarf galaxies, it all started with Sculptor and Fornax, the very first dwarf spheroidal galaxies discovered (\citealt{shapley38a}, \citealt{shapley38b}) to the most recent discoveries (\citealt{willman05b}, \citealt{zucker06}, \citealt{belokurov06}, \citealt{laevens15}, \citealt{drlica-wagner15}, \citealt{kim15b}, \citealt{homma16}, \citealt{torrealba18}, \citealt{homma19}). Their observed diversity in size, luminosity and mass is not a surprise as it naturally arises in various simulations trying to reproduce our Local Universe (\citealt{springel08}, \citealt{vogelsberger14}, \citet{eagle15}, \citealt{wheeler19}). For the Milky Way's globular clusters, \citet{harris10} has compiled the properties of 157 systems and summarized years of efforts of the community to characterize and understand them. That list will only grow longer in the future considering the recent discoveries of new clusters (\citealt{martin16c}, \citealt{koposov17}, \citealt{mau19}), especially with the advent of the Legacy Survey of Space and Time (LSST) in the incoming years (\citealt{ivezic08}, \citealt{simon19}) \\

As increasingly faint MW satellites are discovered, the difficulties in classifying them have grown. Systems fainter than $ M_V\sim -6$ mag are often challenging to classify as globular cluster or galaxy. This issue was notably addressed by \citet{cote02} and \citet{gilmore07} and is commonly known as the ``valley of ambiguity". Distinguishing faint satellite galaxies from globular clusters relies on two main diagnostics: (i) evidence for the presence of substantial amounts of dark matter, and (ii) the presence of a substantial dispersion in metallicity, a feature usually associated with recurrent star formation.  This requires combining deep photometry with extensive spectroscopic campaigns. Dynamical evidence for an excess of mass over that of the stellar component, would then be attributed, in the standard cosmological model, to the presence of a massive DM halo \citep{willman_strader12} and would favour a dwarf galaxy scenario The first measurement of the velocity dispersion of a non-classical MW dwarf galaxy, undertaken by \citet{kleyna05}, showed that Ursa Major~I is indeed DM-dominated. This effort was repeated for most systems known at the time, enriching our knowledge of the MW satellites' properties (\citealt{martin07}, \citealt{simon07}, \citealt{koposov11}, \citealt{simon11}, \citealt{martin16_tri}), an effort that would also rely on other tracers such as the metallicity and the metallicity dispersion of these faint satellites (\citealt{kirby08}, \citealt{koposov15}, \citealt{walker16}). However, in practice, the velocity and metallicity dispersions can be challenging to constrain due to the low number of member stars observed with spectroscopy for a significant fraction of the known faint MW satellites. This problem, that first arose with the study of the kinematics of Segue~2 \citep{kirby13a} still limits our understanding of some of them. The examples over the last few years are numerous, with Draco~II (\citealt{martin16_dra}, \citealt{longeard18}), Tucana~III \citep{simon17}, Colomba~I or Horologium~II \citep{fritz19} that are still not well understood, despite extensive studies with photometry and spectroscopy.

In that context, we present a new spectroscopic study of the faint MW satellite Sagittarius~II (Sgr~II) discovered by \citet{laevens15}, and studied in depth by \citet[L20]{longeard20}. L20 used deep photometry, multi-object spectroscopy and the metallicity-sensitive, narrow-band photometry provided by the Pristine survey \citep{starkenburg17} to refine the structural properties of Sgr~II and infer its dynamical and metallicity properties for the first time. Its velocity dispersion ($\sigma^\mathrm{L20}_\mathrm{vr} = 2.7^{+1.3}_{-1.0}$ km s$^{-1}$) favoured the existence of a low-mass DM halo and therefore a dwarf galaxy scenario. The metallicity of the satellite was found to be in good agreement with the metallicity-luminosity relation of dwarf galaxies of \citet{kirby13}, while its metallicity dispersion was deemed small ($\sigma^\mathrm{L20}_\mathrm{[Fe/H]} = 0.10^{+0.06}_{-0.04}$) but resolved. This last result, however, could be inflated by systematics, an underestimation of the individual uncertainties on the spectroscopic metallicities, or just by the small sample of six stars used to perform the metallicity analysis. If L20 marginally favour Sgr~II to be a very low-mass galaxy, an independent study of the satellite presented in an American Astronomical Society meeting \citep{fu_simon_19}, yet to be published, unequivocally found that Sgr~II is a globular cluster, based on an extremely low metallicity dispersion measurement from their own spectroscopic sample ($< 0.08$ at the 95\% confidence level). Furthermore, if the satellite was found to be physically compact by L20 (r$^\mathrm{L20}_h = 35.5 ^{+1.4}_{-1.2}$ pc), but still possibly within the realm of dwarf galaxies, an estimation of the heliocentric distance of the satellite based on the identification of five RR Lyrae stars potentially members of Sgr~II \citep[V20]{vivas20} would place the system closer than the estimate of L20 ($m-M = 18.97 \pm 0.20$ vs. $19.32^{+0.03}_{-0.02}$ mag, i.e. a difference of $\sim 10$ kpc). This would result in a slight overestimation of its physical size (r$^\mathrm{V20}_h = 30.7 ^{+2.7}_{-2.9}$ pc) and bring the system closer to the realm of old, metal-poor globular clusters. For the rest of this paper, the results of L20 will be used. In many ways, Sgr~II is shrouded in mystery, and the following work will attempt to close the case on the nature of this faint and elusive satellite.

\section{Data selection and acquisition}

\begin{figure}
\begin{center}
\centerline{\includegraphics[width=\hsize]{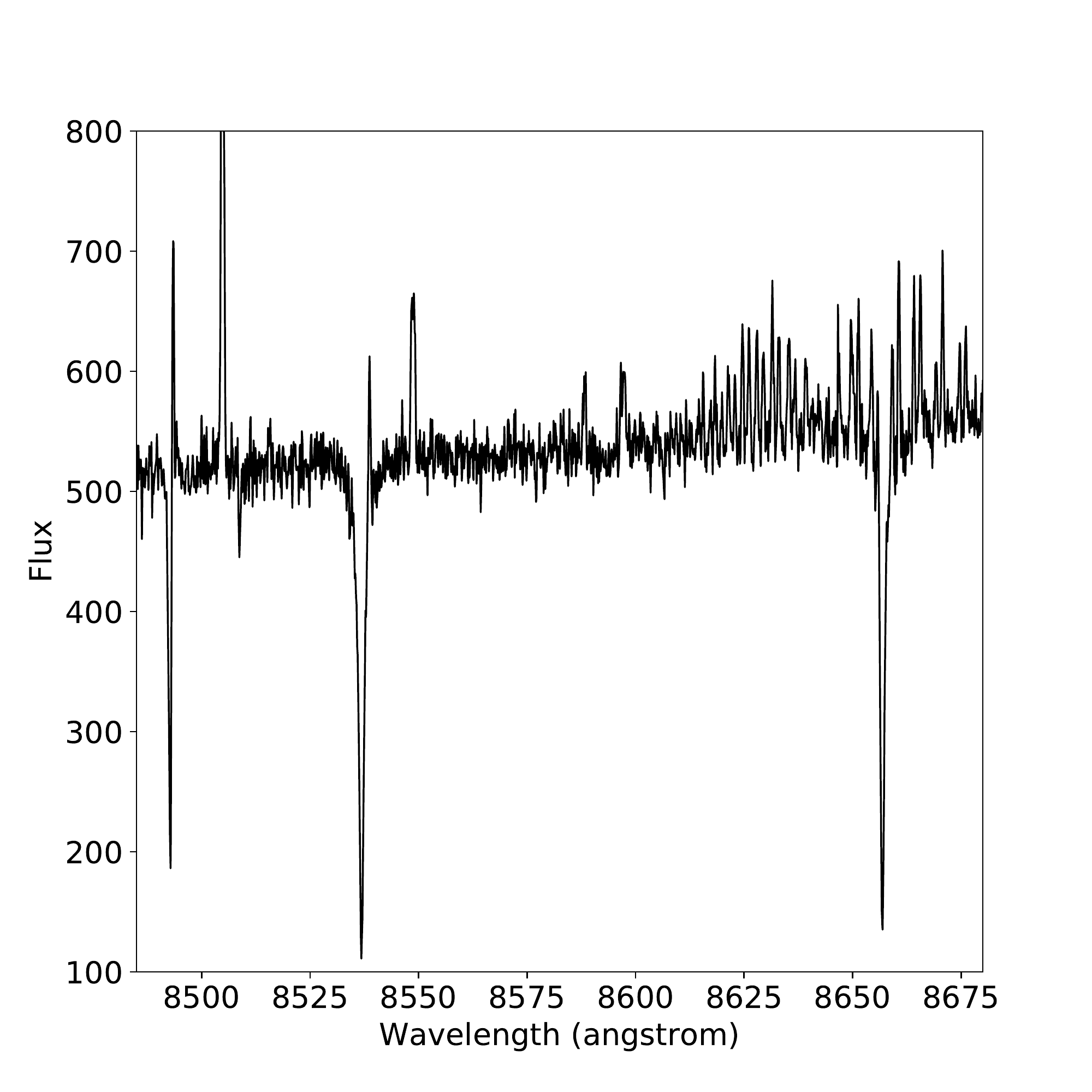}}
\caption{Non sky-substracted spectrum of a Sgr~II member star with a S/N of $\sim 65$ focusing on the Calcium II triplet lines. This star has a metallicity of [Fe/H] $= -2.32 \pm 0.18$ using the calibration of \citet{starkenburg10} as detailed in section 3.2.}
\label{spectrum} 
\end{center}
\end{figure}

\begin{figure*}
\begin{center}
\centerline{\includegraphics[width=\hsize]{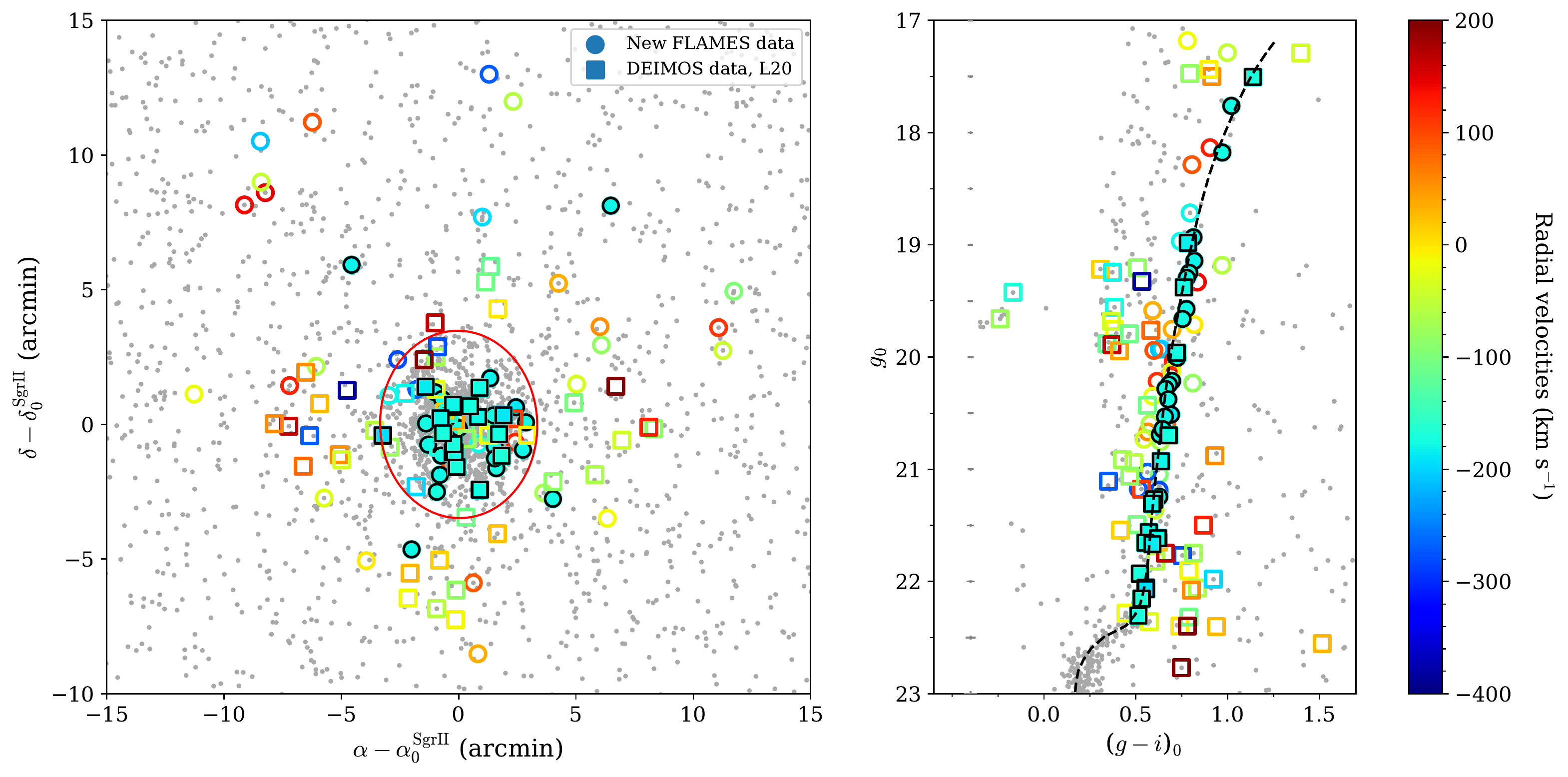}}
\caption{{\textit{Left panel: }}  Spatial distribution of Sgr~II-like stars, i.e. stars with a CMD probability membership of 1 per cent or higher. The field is centered on ($\alpha_0 = 298.16628^\circ, \delta_0 = -22.89633^\circ$). The red contour defines two half-light radii ($r_h \sim 1.7'$) of the satellite, as inferred by L20. The locations of all stars in the spectroscopic dataset are indicated as large, coloured markers. Squares represent the DEIMOS observations from L20, while the circles stand for the new FLAMES data presented in this work. All stars observed spectroscopically are colour-coded according to their heliocentric velocities. Cyan stars have a radial velocity corresponding to Sgr~II at around $-177$ km s$^{-1}$. Filled squares and circles are the member stars of the satellite. {\textit{Right panel: }} CMD of Sgr~II within two half-light radii (grey) superimposed with the entire spectroscopic dataset. The best-fitting Darmouth isochrone from L20 ($12$ Gyr, [Fe/H] $= -2.35$, $[\alpha$/Fe] $= 0$, $m-M = 19.32$ mag) is shown as a black dashed line and is perfectly compatible with the identified members of Sgr~II.}
\label{cmd} 
\end{center}
\end{figure*}

\begin{figure}
\begin{center}
\centerline{\includegraphics[width=\hsize]{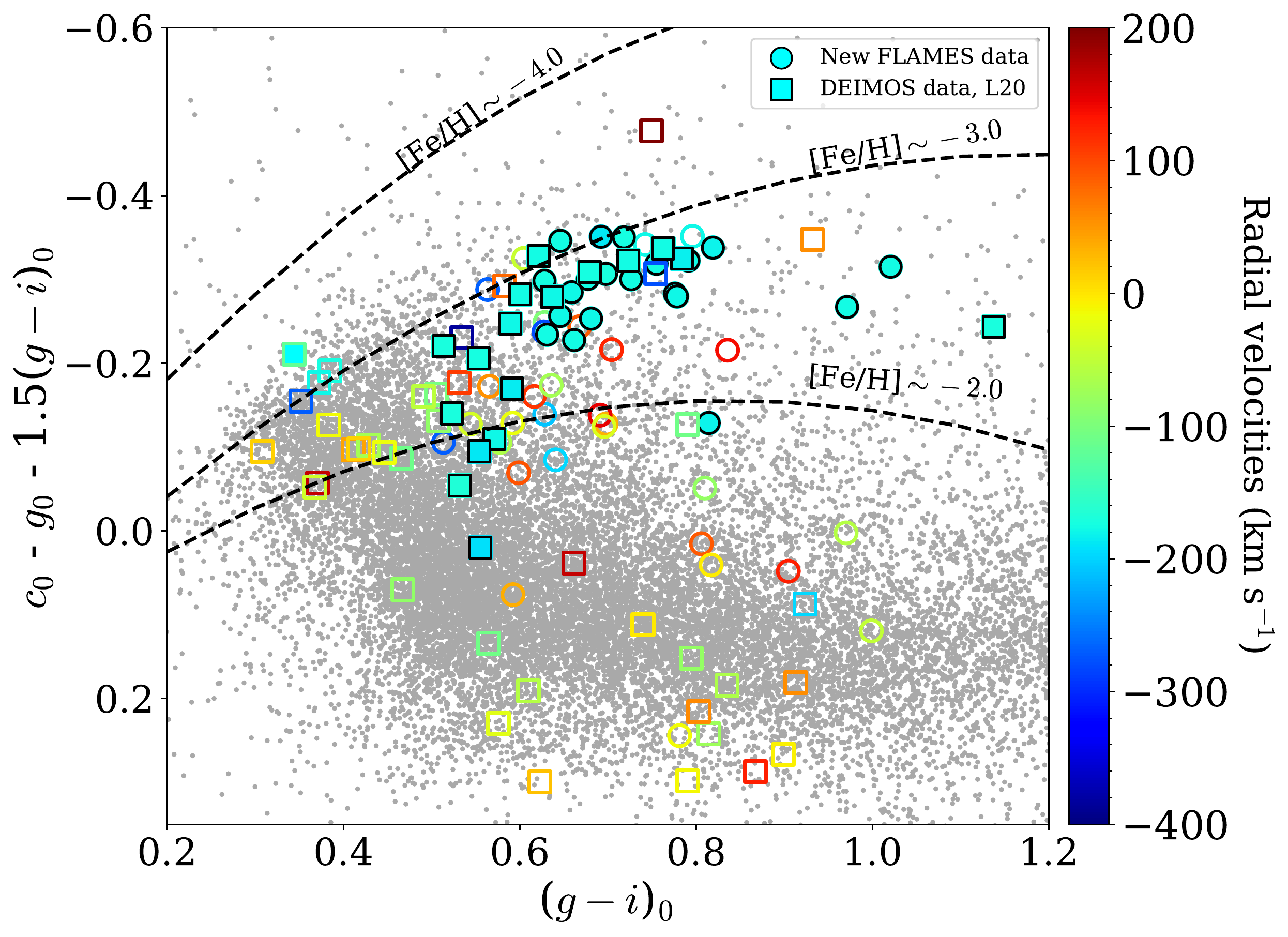}}
\caption{Pristine colour-colour diagram of Sgr~II showing field stars (grey) and the spectroscopic dataset with large, coloured markers. Member stars of Sgr~II are shown with filled markers. While the x-axis is a temperature proxy, the y-axis contains the Pristine metallicity-sensitive, narrow-band photometry denoted $c_0$, and therefore the metallicity information. As a result, stars are distributed according to their metallicity: the ones around solar metallicity form the stellar locus where most of the grey field stars are located. As a star goes upwards in the diagram (i.e. lower y-axis value) for a fixed temperature, its metallicity decreases. This is represented more visually with the three iso-metallicity sequences in black dashed lines, showing where stars with a [Fe/H] of $-4.0$, $-3.0$ and $-2.0$ should be located. This plot shows the efficiency of Pristine in identifying Sgr~II members, as almost all DEIMOS stars dynamically compatible with the satellite come out as metal-poor.}
\label{cahk} 
\end{center}
\end{figure}

The spectroscopy used in this work is a combination of two datasets. The first data were observed with the DEep Imaging Multi-Object Spectrograph (DEIMOS, \citealt{faber03}), already detailed and analysed in L20. The second dataset is composed of new observations performed with the Fibre Large Array Multi Element Spectrograph \citep[FLAMES]{pasquini02} mounted on the Very Large Telescope (VLT). The \mbox{GIRAFFE}/HR21 grating, with a resolution of R $\sim 18000$, a central wavelength of $8757$ \AA $\; $and a bandwidth between 8484 and 9001\AA$\;$was used to resolve the Calcium II triplet infrared lines. These FLAMES observations were conducted during three different nights to attempt to detect binary stars: on the 22.06.16 ($1 \times 2610$s of integration), the 21.07.16 ($2 \times 2610$s) and the 30.07.16 ($2 \times 2610$s).  

The data were reduced by the European Southern Observatory team using their pipeline specifically tailored to FLAMES data \citep{melo09}. The spectrum of a Sgr~II member star is displayed in Figure \ref{spectrum} to illustrate the quality of the data. In order to infer the radial velocities, the equivalent widths and their respective uncertainties, the method already used in L20 was used and consists of determining the combination of Gaussian profiles that best fits a continuum composed of the Calcium II triplet lines at rest as the unique spectral features. It also includes a correction to account for the non-gaussianity of the wings' lines. The typical uncertainty on the radial velocities for the FLAMES dataset is of the order of $1$ km s$^{-1}$, and all the stars with signal-to-noise (S/N) ratio below 3 per resolution element are discarded from the sample. In the same fashion as in \citet{simon_geha07} and L20, the systematic error on the velocities is determined from stars observed more than once in the FLAMES sample. The resulting systematics are a bias of $ b = 0.4 \pm 0.3$ km s$^{-1}$ and a standard deviation of $ \delta_\mathrm{thr} = 0.8 \pm 0.1$ km s$^{-1}$. \\

To build our list of FLAMES targets, we used the locations of stars in the colour-magnitude diagram (CMD) of Sgr II built from deep MegaCam photometry and shown in the right panel of Figure \ref{cmd}, along with their spatial distribution in the left panel. More importantly, this spectroscopic study benefits from the use of the narrow-band, metallicity-sensitive photometry provided by the Pristine survey \citep{starkenburg17}.  All stars observed in Pristine have a photometric metallicity measurement, as illustrated in the colour-colour diagram shown in Figure \ref{cahk}. In this diagram, stars are distributed according to their metallicities, from the most metal-rich at the bottom of the plot, to the most metal-poor at the top, down to a metallicity of [Fe/H] $\sim -4.0$. Using the prior knowledge that Sgr~II is indeed a very metal-poor satellite ([Fe/H]$_\mathrm{SgrII} \sim -2.28$, L20), the target list was built so that the stars observed are confirmed to be very metal-poor according to the Pristine survey. In doing so, we increase the probability of finding new members (\citealt{youakim17}, \citealt{aguado19}). The DEIMOS spectroscopy supplementing this dataset consists only of the DEIMOS stars identified as non variables, non binaries and metal-poor according to Pristine ([Fe/H]$_\mathrm{CaHK} < -1.6$) by L20.\\

In order to clean the FLAMES data, we check for the existence of binaries in the sample. For a given star, each measurement from the $k$-th epoch is assumed to be reasonably well-described by a Gaussian distribution centered on the radial velocity measurement and a standard deviation corresponding to its uncertainty. Then, the probability that the measurement $k$ is discrepant from that of the same star at another epoch $j$ at the $2\sigma$ level is computed. This probability corresponds to the star being variable between epochs $k$ and $j$. The same procedure is used to compute the probability that each velocity measurement of a given epoch is compatible with all the others within $2\sigma$, i.e. the probability that it is not varying. If the product of the probabilities representing the hypothesis "variable" is greater than the hypothesis "non-variable", then the star is considered a binary star and discarded from the main sample so that their variability does not affect the dynamical analysis of Sgr~II. We find a total of 6 potential binaries in the sample, however, none have the right velocity to be a member of Sgr~II. 

The final spectroscopic sample consists of 113 stars, with 47 new stars observed with FLAMES as detailed in Table \ref{table1} with four stars out of those 47 in common with L20's sample. As a consistency check, we compare the velocities obtained with FLAMES and DEIMOS for these four stars and find no statistical differences between the two samples.

\section{Results}

In this section, we constrain the systemic heliocentric velocity of Sgr~II and its associated velocity dispersion.

\begin{figure}
\begin{center}
\centerline{\includegraphics[width=\hsize]{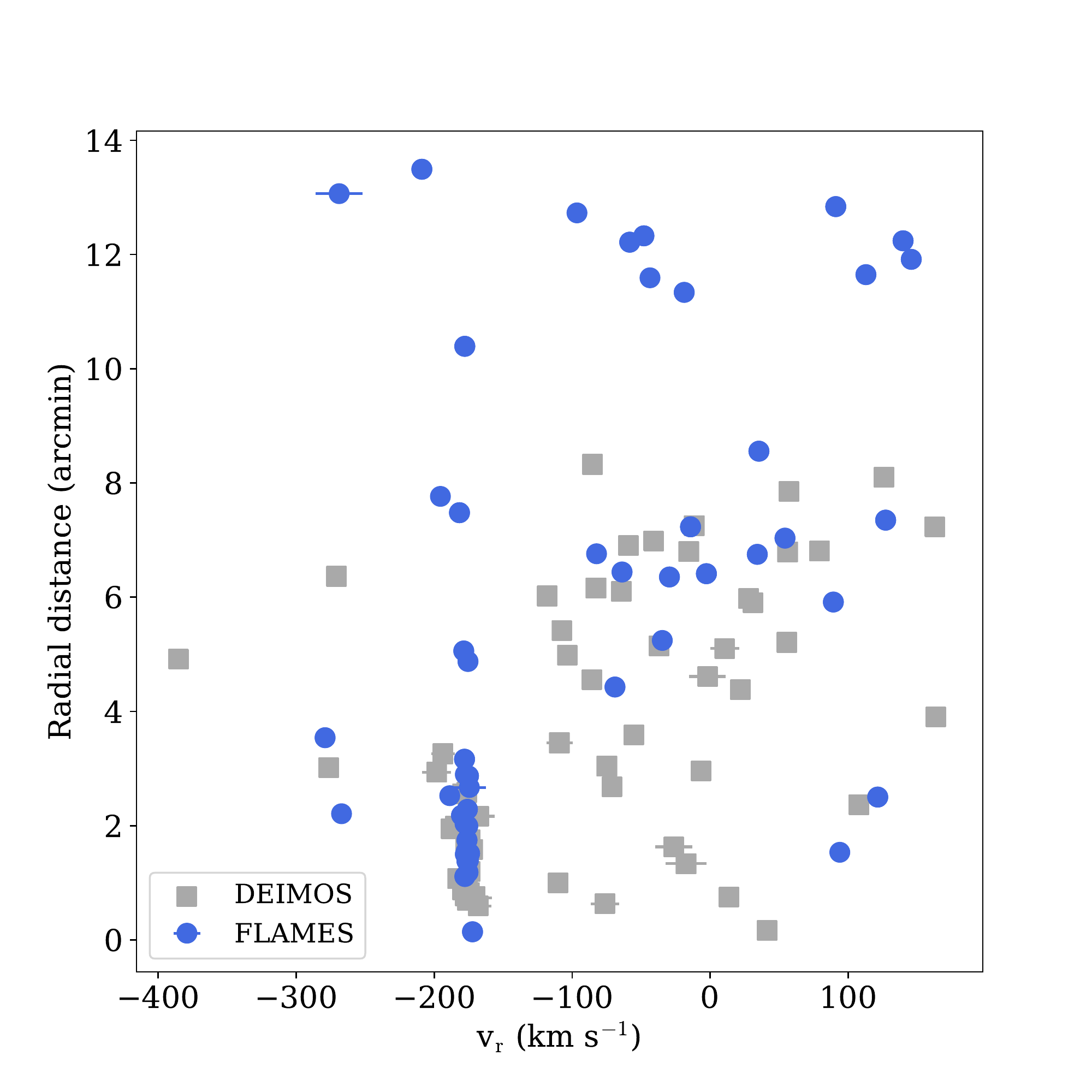}}
\caption{Heliocentric velocities vs. radial distance to Sgr~II's centroid for FLAMES (blue circles) and DEIMOS (grey squares) datasets. The addition of the FLAMES data doubles the number of identified members for Sgr~II. All error bars not appearing in this plot are smaller than the size of the circles/squares.}
\label{histos_vel} 
\end{center}
\end{figure}

\subsection{Dynamical properties}

\begin{figure}
\begin{center}
\centerline{\includegraphics[width=\hsize]{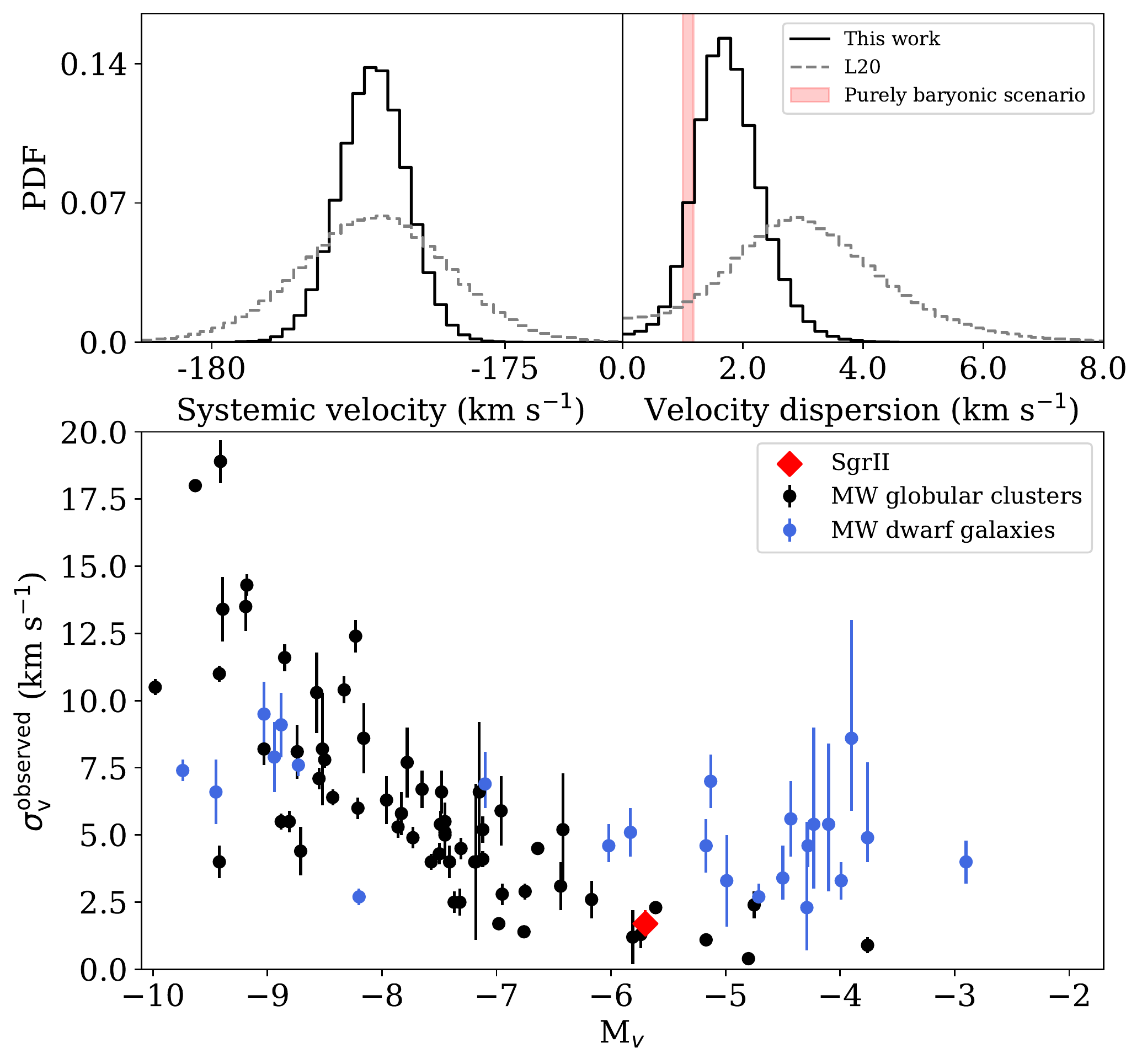}}
\caption{1D marginalised PDFs of the systemic radial velocity (top left panel) and velocity dispersion (top right panel) of Sgr~II from this work as black solid lines, and L20 in grey dashed lines using only DEIMOS data. The expected velocity dispersion for a Sgr~II-like typical MW globular cluster ($1.1 \pm 0.1$ km s$^{-1}$) is indicated as the red shaded region. The bottom panel shows the observed velocity dispersions of most globular clusters (black) and confirmed dwarf galaxies (blue) of the MW as a function of their absolute magnitudes. The properties of the globular clusters were taken from the \citet{harris10} catalog and references therein. The dwarf galaxy measurements come from \citet{simon19} and references therein. In this plot, the location of Sgr~II is indicated by the red diamond.}
\label{pdfs_vel} 
\end{center}
\end{figure}

All stars in our spectroscopic sample with a heliocentric velocity are shown in Figure \ref{histos_vel}, along with a more detailed view. As already detailed in section 2, the final sample consists of all non-binary, non-HB stars compatible with Sgr~II's sequence in the CMD and that appear as very metal-poor using the CaHK photometry. Stars with mediocre CaHK photometry from the DEIMOS catalog are also included as their photometric metallicities are not reliable. Thanks to the photometry provided by the Pristine survey, the FLAMES data are already quite clean with only a few clear contaminants discrepant from the Sgr~II population at around $-180$ km s$^{-1}$. When available, the proper motions provided by the Gaia Data Release 2 \citep{brown18} are also used to filter out contaminants. This procedure allows us to get rid of one star with CMD and metallicity properties strongly compatible with Sgr~II, but a proper motion unequivocally discrepant from that of the satellite (\citealt{massari18}, L20).

In order to derive the dynamical properties of Sgr~II, the velocity distribution of our dataset and shown in Figure \ref{histos_vel} is assumed to be the sum of two Gaussian distributions, one standing for Sgr~II's population and the other for the MW contamination. The favoured dynamical model is obtained through a Monte Carlo Markov Chain \citep[MCMC]{hastings70} algorithm. \\

The Probability Distribution Functions (PDFs) of the systemic heliocentric velocity and the velocity dispersion of Sgr~II are displayed in the top panels of Figure \ref{pdfs_vel}. The resulting systemic velocity $\langle \mathrm{v}_\mathrm{SgrII} \rangle$ is $-177.2_{-0.6}^{+0.5}$ km s$^{-1}$ and is perfectly compatible with the one of L20. The velocity dispersion found in this work is also compatible with L20 ($\sigma^\mathrm{SgrII}_\mathrm{v} = 2.7_{-1.0}^{+1.3}$ km s$^{-1}$), but with much tighter constraints, it is measured to be $\sigma^\mathrm{SgrII}_\mathrm{v} = 1.7 \pm 0.5$ km s$^{-1}$. Using the formalism of \citet{wolf10} which assumes dynamical equilibrium and a flat velocity dispersion profile, this results into a mass-to-light (M/L) ratio of $3.5^{+2.7}_{-1.6}$ M$_{\odot}$ L$_{\odot}^{-1}$. Sgr~II shows no sign of a velocity dispersion gradient as the determination of its dynamical properties by selecting stars inside and outside 1 arcminute yields no statistical difference. Furthermore, adding another contaminating population to the likelihood model does not change our results.

We classify as a member of Sgr~II any star with  line-of-sight velocity and proper motion membership probabilities (when available) above 50\%. We identify $22$ member stars in the FLAMES data. 3 of them were already identified in the L20 dataset, leading to the identification of 19 new members. Combined with the spectroscopically confirmed members of L20, 43 stars are confirmed to belong to Sgr~II, including binaries and horizontal branch (HB) stars. This underlines the importance of the Pristine photometry. To estimate the success rate of the FLAMES data in identifying new members, we do not simply take the fraction of confirmed members over the overall number of stars observed since the sample of Sgr~II candidates identified before observation was not large enough to fill all the FLAMES fibers. Among the 47 new stars observed, only $32$ were considered as promising Sgr~II candidates. Therefore, it yields a success rate of $\sim 60 \%$ for the FLAMES data sample based on the Pristine selection. In comparison, the DEIMOS dataset of L20, solely based on a CMD-based selection, had a 20\% success rate. The contrast is even more striking when considering that the DEIMOS selection focuses on the central region of the satellite and is therefore more likely to find Sgr~II stars, while the FLAMES observations had a much wider field of view and aim to find stars located much further away than Sgr~II's centroid.

\subsection{Metallicity properties} 

\begin{figure}
\begin{center}
%\centerline{\includegraphics[width=\hsize]{FeH_spectro_histo_FLAMES+DEIMOS.pdf}}
\centerline{\includegraphics[width=\hsize]{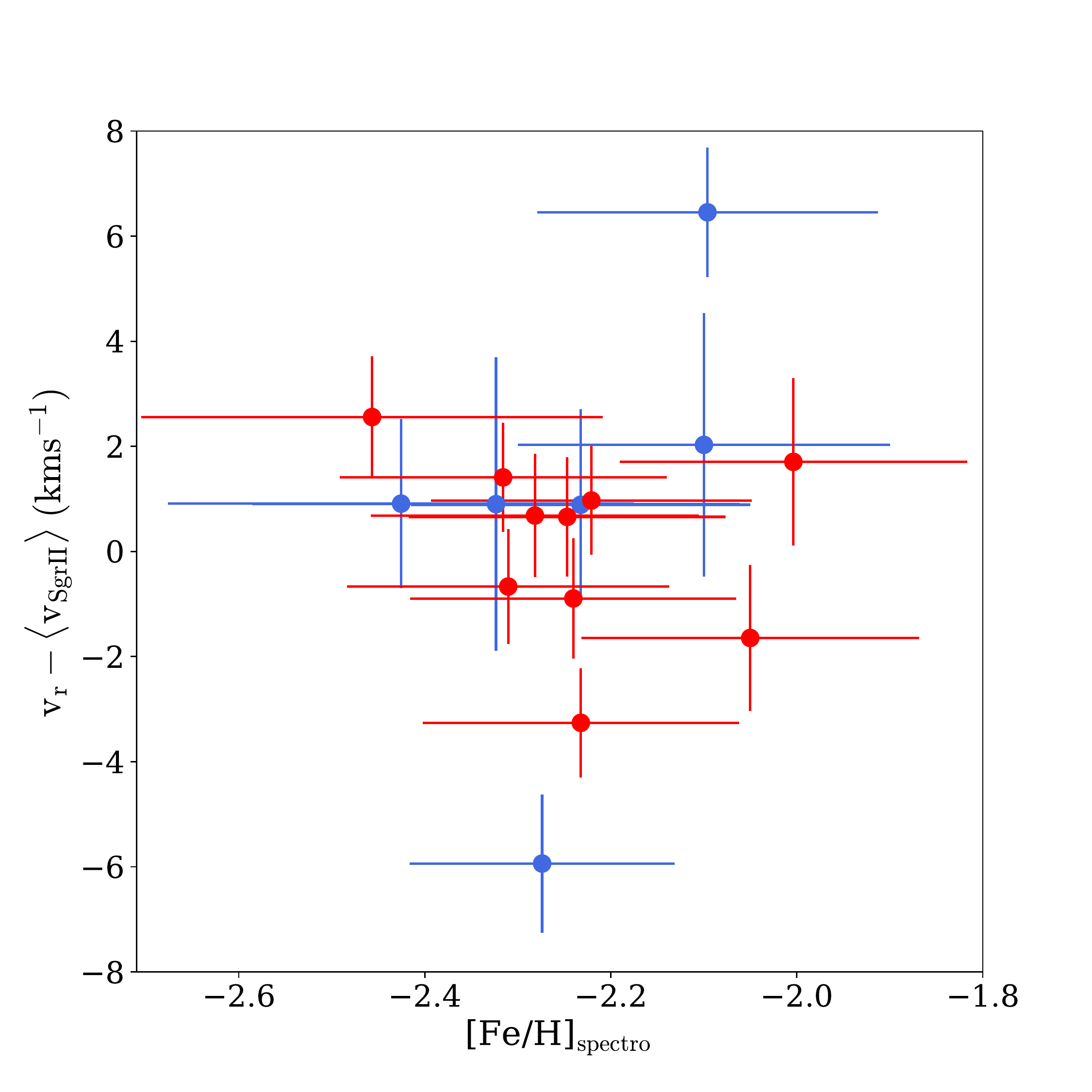}}
\caption{Individual metallicities of all bright stars in the spectroscopic sample used to derive the metallicity properties of Sgr~II using the calibration of \citet{starkenburg10}. Each measurement is modelled by a Gaussian with a mean corresponding to the favoured metallicity of the star, and a standard deviation equal to its uncertainty. Stars from the L20 sample are shown as blue dots, while the new FLAMES stars are represented in red.}
\label{pdfs_feh} 
\end{center}
\end{figure}

\begin{figure}
\begin{center}
\centerline{\includegraphics[width=\hsize]{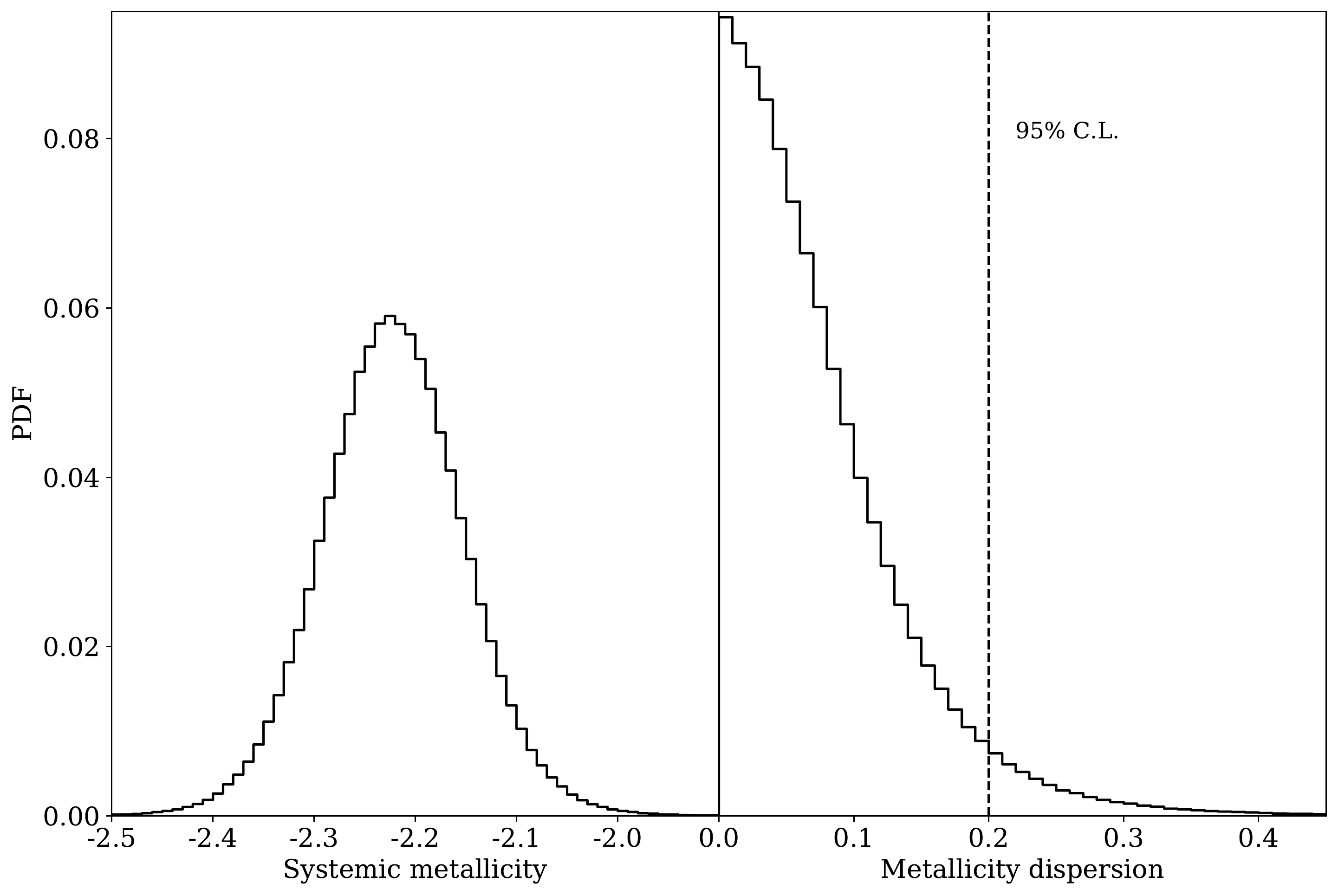}}
\caption{One-dimensional PDFs of the systemic metallicity (left panel) and metallicity dispersion (right panel) of Sgr~II with spectroscopy. The 95\% C.L. on the metallicity dipersion is hown as a black dashed line. }
\label{metallicities} 
\end{center}
\end{figure}

In the following section, the mean stellar metallicity of Sgr~II and its metallicity dispersion are derived using our new FLAMES data. To do so, we select only non-binary, non-HB stars with a spectral signal-to-noise ratio greater than 12 and with $g_0 < 20.5$ (i.e. $\sim$ 1 mag below the HB of Sgr~II). This yields a subsample of 17 stars to extract a measurement of the spectroscopic metallicity. This procedure is performed through the use of the empirical calibration of \citet{starkenburg10} that uses the equivalent widths (EWs) of the Calcium II triplet lines to deduce a measurement of the metallicity of a star. The uncertainties on the coefficients defining the polynomials to transform a set of EWs into a [Fe/H]$_\mathrm{spectro}$ are folded in with the uncertainties on the EWs through a Monte Carlo procedure to obtain the final uncertainties $\delta_\mathrm{[Fe/H]}$ on each individual spectroscopic metallicity measurement. We take the value of 8\% reported by \citet{starkenburg10} to be the uncertainty on each coefficients. We note that these uncertainties do not include several systematics (due to, for instance, the continuum placement, or the detailed chemical abundance pattern of the star when transforming a Ca II measurement into [Fe/H]). Figure \ref{pdfs_feh} shows the spectroscopic metallicities of the 17 stars and their uncertainties assuming each measurement can be modelled by a Gaussian centered on [Fe/H]$_\mathrm{spectro}$ and with a standard deviation corresponding to $\delta_\mathrm{[Fe/H]}$. For stars observed more than once, a metallicity measurement is derived for each epoch and we verify that each epoch yields a metallicity compatible with the others. This procedure highlights two stars with discrepant metallicity measurements at different epochs. We decide to discard those stars for the rest of the analysis, however, including them has a very limited impact on the results. None of the 4 stars in common between the FLAMES and DEIMOS samples have a spectroscopic metallicity measurement.

Since all 15 stars in the subsample are likely Sgr~II members, their metallicity distribution is assumed to be only reflective of Sgr~II population, which is therefore modelled with a Gaussian distribution weighted with the membership probability of each star according to L20. We find a spectroscopic systemic metallicity of $\langle$[Fe/H]$^\mathrm{SgrII}_\mathrm{spectro} \rangle = -2.23 \pm 0.06$, and an unresolved metallicity dispersion with $\sigma^\mathrm{SgrII}_\mathrm{spectro} < 0.20$ at the 95\% confidence limit. The PDFs corresponding to these results are shown in Figure \ref{metallicities}. The metallicity dispersion is unresolved, and is constrained to be below 0.20 dex at the 95\% confidence limit (C.L.).

\section{Discussion}
Throughout this paper, we analyse the dynamical and metallicity properties of the faint MW satellite Sgr~II. To this end, we combine the Keck II/DEIMOS spectroscopic dataset of our previous analysis of the satellite with new VLT/FLAMES data. These FLAMES observations are the first ones to be carried out by selecting \textit{a priori} the interesting candidates using the narrow-band, metallicity sensitive photometry of the Pristine survey for a faint satellite of the MW.  The spectroscopic observations presented in this work are decisive as they double the overall sample of members stars known. Nine new stars are bright enough to estimate their spectroscopic metallicities with the empirical Calcium II triplet calibration of \citet{starkenburg10}, while the sample of L20 only consists of 6 stars. Therefore, this study considerably enlarges the statistics available for the satellite and allows us to put much tighter constraints on the dynamical and metallicity properties of SgrII. 

The systemic radial velocity of Sgr~II is found to be $\langle \mathrm{v}_\mathrm{SgrII} \rangle = -177.2_{-0.6}^{+0.5}$ km s$^{-1}$. The velocity dispersion of $\sigma_\mathrm{v}^\mathrm{SgrII} = 1.7 \pm 0.5$ km s$^{-1}$ is well among the range observed for the MW globular clusters at the same luminosity, as shown in Figure \ref{pdfs_vel}. It translates into a M/L ratio of $3.5^{+2.7}_{-1.6}$ M$_{\odot}$ L$^{-1}_{\odot}$, thus suggesting that the dynamics of the satellite is not mainly driven by a DM halo. We identify 19 new members in the faint system, for a total of 43 confirmed spectroscopic Sgr~II members. With a fraction of member identified over the entirety of the FLAMES sample of $60$\%, the \textit{a priori} selection using Pristine performs three times better than the simple CMD-based selection of L20, despite a riskier observational strategy through the search for stars located beyond two half-light radii of the satellite. Four of these stars are identified, with the outermost one lying at $6.2$r$_{h}$. 15 out of the 43 members have good enough quality spectra to measure the EWs of their Calcium II triplet and measure their metallicity using the empirical calibration of \citet{starkenburg10}. We find a systemic metallicity of $-2.23 \pm 0.07$ and constrain the metallicity dispersion of the satellite to be less than $0.20$ at the $95$\% confidence limit.

 The velocity dispersion as inferred by L20 is larger than the one found in the present analysis. Furthermore, L20's metallicity analysis indicates that there are multiple stellar populations in the system, as the metallicity dispersion is resolved both with CaHK photometry and spectroscopy. This is not the case anymore in this work as the spectroscopic metallicity distribution of our dataset is well-described with only one stellar population despite the addition of 9 new stars with spectroscopic metallicites. Moreover, Figure \ref{pdfs_feh} shows that no star with a metallicity below $-2.5$ is confidently detected, a limit commonly attributed to the lowest metallicity achievable by globular clusters (\citealt{harris10} and references therein, \citealt{beasley19}). Therefore, we are able to conclude that Sgr~II is a globular cluster. 
 
Nonetheless, the satellite is still quite extended when compared to other MW globular clusters of the same luminosity. Within a range of one magnitude around Sgr~II's absolute magnitude, the largest MW cluster is Pal~5 (M$_V \sim -5.2 $), a fairly metal-poor system ([Fe/H] $\sim -1.4$) with a size of $\sim 19.2$ pc, while most of the others have a size below $10$ pc \citep{harris10}. Whether we consider the distance of L20 or \citet{vivas20}, Sgr~II is still at least 1.5 times more extended than Pal~5. Sgr~II is a case in point of the ambiguity surrounding the faint MW satellites discovered in recent years, with a somewhat larger half-light radius than expected from GCs at that luminosity, a metallicity perfectly compatible with the luminosity-metallicity relation of dwarf galaxies \citep{kirby13}, and velocity and metallicity dispersions challenging to resolve. Though we conclude that Sgr~II is an old and metal-poor globular cluster, the system remains interesting in many aspects, from its possible association to the Sgr stream to an even deeper understanding of the unusually large cluster, which could build a bridge to the still exclusive definitions of clusters and galaxies of the MW.

\section*{Acknowledgments}
We gratefully thank the CFHT staff for performing the observations in queue mode, for their reactivity in adapting the schedule, and for answering our questions during the data-reduction process.

NFM, RI, and NL gratefully acknowledge support from the French National Research Agency (ANR) funded project ``Pristine'' (ANR-18-CE31-0017) along with funding from CNRS/INSU through the Programme National Galaxies et Cosmologie and through the CNRS grant PICS07708. ES gratefully acknowledges funding by the Emmy Noether program from the Deutsche Forschungsgemeinschaft (DFG). This work has been published under the framework of the IdEx Unistra and benefits from a funding from the state managed by the French National Research Agency as part of the investments for the future program. The authors thank the International Space Science Institute, Berne, Switzerland for providing financial support and meeting facilities to the international team ``Pristine''.

Based on observations collected at the European Southern Observatory under ESO programme(s) 099.B-0690(A).

Some of the data presented herein were obtained at the W. M. Keck Observatory, which is operated as a scientific partnership among the California Institute of Technology, the University of California and the National Aeronautics and Space Administration. The Observatory was made pos- sible by the generous financial support of the W. M. Keck Foundation. Furthermore, the authors wish to recognize and acknowledge the very significant cultural role and rever- ence that the summit of Maunakea has always had within the indigenous Hawaiian community. We are most fortunate to have the opportunity to conduct observations from this mountain.
 
Based on observations obtained at the Canada-France-Hawaii Telescope (CFHT) which is operated by the National Research Council of Canada, the Institut National des Sciences de l'Univers of the Centre National de la Recherche Scientifique of France, and the University of Hawaii.

This work has made use of data from the European Space Agency (ESA)
mission {\it Gaia} (\url{https://www.cosmos.esa.int/gaia}), processed by
the {\it Gaia} Data Processing and Analysis Consortium (DPAC,
\url{https://www.cosmos.esa.int/web/gaia/dpac/consortium}). Funding
for the DPAC has been provided by national institutions, in particular
the institutions participating in the {\it Gaia} Multilateral Agreement.

\clearpage

\newpage

\begin{table*}

\caption{Properties of the new FLAMES spectroscopic sample. The Pristine metallicity of a given star is indicated only if [Fe/H]$_\mathrm{CaHK} < -1.0$. The individual spectroscopic metallicity is reported for stars with S/N $>= 12$ and $g_0 <= 20.5$ mag. The systematic threshold $\delta_{\mathrm{thr}}$ is not included in the velocity uncertainties presented in this table.
\label{table1}}

\setlength{\tabcolsep}{2.5pt}
\renewcommand{\arraystretch}{0.3}
\begin{sideways}
\begin{tabular}{cccccccccccccc}
\hline
RA (deg) & DEC (deg) & $g_0$ & $i_0$ & $CaHK_0$ & $\mathrm{v}_{r} (\kms)$ & $\mu_{\alpha}^{*}$ (mas.yr$^{-1}$) & $\mu_{\delta}$ (mas.yr$^{-1}$) &  S/N & [Fe/H]$_{\mathrm{CaHK}}$ & [Fe/H]$^\mathrm{S10}_\mathrm{spectro}$ & Member\\
\hline

298.37718 & $-$21.98293 & 21.05 $\pm$ 0.01 & 20.42 $\pm$ 0.01 & 21.74 $\pm$ 0.04 & -96.5 $\pm$ 1.4 & 5.0 $\pm$ 4.0 & -5.1 $\pm$ 2.1 & 3.5 & $-$2.50 $\pm$ 0.12 & --- &  N  \\ \\ 
298.28281 & $-$21.92996 & 19.57 $\pm$ 0.01 & 18.80 $\pm$ 0.01 & 20.45 $\pm$ 0.02 & -177.8 $\pm$ 0.6 & -0.2 $\pm$ 0.5 & -1.1 $\pm$ 0.3 & 24.4 & $-$2.58 $\pm$ 0.12 & -2.2 $\pm$ 0.11 &  Y  \\ \\ 
298.24280 & $-$21.97804 & 19.75 $\pm$ 0.01 & 19.05 $\pm$ 0.01 & 20.67 $\pm$ 0.02 & 34.1 $\pm$ 2.4 & -1.9 $\pm$ 2.7 & -1.9 $\pm$ 1.3 & 12.7 & $-$1.90 $\pm$ 0.12 & -1.5 $\pm$ 0.21 &  N  \\ \\ 
298.20794 & $-$21.86545 & 19.18 $\pm$ 0.01 & 18.21 $\pm$ 0.01 & 20.64 $\pm$ 0.02 & -58.3 $\pm$ 0.5 & -0.6 $\pm$ 0.2 & -1.0 $\pm$ 0.1 & 48.9 & $-$1.25 $\pm$ 0.12 & -1.26 $\pm$ 0.16 &  N  \\ \\ 
298.18438 & $-$21.93702 & 19.93 $\pm$ 0.01 & 19.29 $\pm$ 0.01 & 20.80 $\pm$ 0.02 & -195.5 $\pm$ 3.3 & --- $\pm$ --- & --- $\pm$ --- & 6.3 & $-$1.72 $\pm$ 0.12 & --- &  N  \\ \\ 
298.18958 & $-$21.84866 & 21.18 $\pm$ 0.01 & 20.66 $\pm$ 0.01 & 21.84 $\pm$ 0.04 & -268.8 $\pm$ 16.9 & 1.7 $\pm$ 0.2 & -2.0 $\pm$ 0.1 & 3.1 & $-$2.01 $\pm$ 0.12 & --- &  N  \\ \\ 
298.05410 & $-$21.87831 & 18.29 $\pm$ 0.01 & 17.48 $\pm$ 0.01 & 19.51 $\pm$ 0.01 & 91.0 $\pm$ 0.3 & --- $\pm$ --- & --- $\pm$ --- & 79.8 & $-$1.34 $\pm$ 0.12 & -1.43 $\pm$ 0.16 &  N  \\ \\ 
298.08415 & $-$21.96661 & 18.93 $\pm$ 0.01 & 18.12 $\pm$ 0.01 & 20.03 $\pm$ 0.01 & -181.7 $\pm$ 0.4 & -12.8 $\pm$ 0.7 & -18.6 $\pm$ 0.4 & 60.6 & $-$1.99 $\pm$ 0.12 & -1.8 $\pm$ 0.13 &  N  \\ \\ 
298.00202 & $-$21.92952 & 19.33 $\pm$ 0.01 & 18.50 $\pm$ 0.01 & 20.37 $\pm$ 0.02 & 139.8 $\pm$ 0.4 & --- $\pm$ --- & --- $\pm$ --- & 31.3 & $-$2.28 $\pm$ 0.12 & -2.12 $\pm$ 0.11 &  N  \\ \\ 
298.01422 & $-$21.89008 & 19.93 $\pm$ 0.01 & 19.30 $\pm$ 0.01 & 20.73 $\pm$ 0.02 & -208.9 $\pm$ 1.0 & -4.7 $\pm$ 0.2 & -6.0 $\pm$ 0.1 & 16.4 & $-$2.00 $\pm$ 0.12 & -2.24 $\pm$ 0.09 &  N  \\ \\ 
298.01803 & $-$21.92193 & 20.17 $\pm$ 0.01 & 19.47 $\pm$ 0.01 & 21.06 $\pm$ 0.03 & 145.7 $\pm$ 0.9 & 4.2 $\pm$ 1.2 & -6.5 $\pm$ 0.9 & 14.6 & $-$1.95 $\pm$ 0.12 & -1.13 $\pm$ 0.22 &  N  \\ \\ 
298.36550 & $-$22.00547 & 20.22 $\pm$ 0.01 & 19.60 $\pm$ 0.01 & 20.98 $\pm$ 0.02 & 112.9 $\pm$ 2.1 & -0.6 $\pm$ 0.6 & -1.4 $\pm$ 0.4 & 15.3 & $-$2.10 $\pm$ 0.12 & -2.36 $\pm$ 0.12 &  N  \\ \\ 
298.36878 & $-$22.01959 & 21.36 $\pm$ 0.01 & 20.76 $\pm$ 0.01 & 21.95 $\pm$ 0.04 & -43.6 $\pm$ 3.0 & 4.8 $\pm$ 3.1 & 2.9 $\pm$ 1.8 & 3.6 & $-$3.11 $\pm$ 0.12 & --- &  N  \\ \\ 
298.27562 & $-$22.01614 & 20.23 $\pm$ 0.01 & 19.42 $\pm$ 0.01 & 21.40 $\pm$ 0.03 & -82.3 $\pm$ 2.8 & -0.6 $\pm$ 1.3 & -11.3 $\pm$ 0.8 & 4.1 & $-$1.58 $\pm$ 0.12 & --- &  N  \\ \\ 
298.27457 & $-$22.00481 & 20.67 $\pm$ 0.01 & 20.10 $\pm$ 0.01 & 21.34 $\pm$ 0.03 & 54.2 $\pm$ 1.5 & --- $\pm$ --- & --- $\pm$ --- & 10.1 & $-$2.23 $\pm$ 0.12 & --- &  N  \\ \\ 
298.23845 & $-$22.11143 & 20.69 $\pm$ 0.01 & 20.06 $\pm$ 0.01 & 21.40 $\pm$ 0.03 & -175.5 $\pm$ 0.9 & -0.6 $\pm$ 1.3 & -0.0 $\pm$ 0.8 & 10.8 & $-$2.41 $\pm$ 0.12 & --- &  Y  \\ \\ 
298.25661 & $-$22.0404 & 20.73 $\pm$ 0.01 & 20.19 $\pm$ 0.01 & 21.42 $\pm$ 0.03 & -34.7 $\pm$ 1.4 & 0.4 $\pm$ 1.1 & 0.0 $\pm$ 0.7 & 6.1 & $-$2.03 $\pm$ 0.12 & --- &  N  \\ \\ 
298.23141 & $-$22.10773 & 20.75 $\pm$ 0.01 & 20.12 $\pm$ 0.01 & 21.53 $\pm$ 0.03 & -69.0 $\pm$ 1.7 & -1.1 $\pm$ 1.1 & -1.8 $\pm$ 0.6 & 4.4 & $-$2.13 $\pm$ 0.12 & --- &  N  \\ \\ 
298.19512 & $-$22.09244 & 18.18 $\pm$ 0.01 & 17.21 $\pm$ 0.01 & 19.37 $\pm$ 0.01 & -175.9 $\pm$ 0.3 & -13.5 $\pm$ 0.6 & -25.9 $\pm$ 0.4 & 95.7 & $-$2.31 $\pm$ 0.12 & -2.15 $\pm$ 0.12 &  Y  \\ \\ 
298.18145 & $-$22.07735 & 18.72 $\pm$ 0.01 & 17.92 $\pm$ 0.01 & 19.56 $\pm$ 0.01 & -177.7 $\pm$ 0.8 & -7.7 $\pm$ 1.3 & -9.1 $\pm$ 0.7 & 18.3 & $-$2.97 $\pm$ 0.12 & -2.18 $\pm$ 0.11 &  N  \\ \\ 
298.11270 & $-$22.04758 & 18.97 $\pm$ 0.01 & 18.23 $\pm$ 0.01 & 19.74 $\pm$ 0.01 & -178.0 $\pm$ 0.9 & -1.3 $\pm$ 1.1 & -8.7 $\pm$ 0.6 & 11.9 & $-$2.91 $\pm$ 0.12 & -2.13 $\pm$ 0.18 &  N  \\ \\ 
298.19043 & $-$22.03676 & 19.14 $\pm$ 0.01 & 18.32 $\pm$ 0.01 & 20.03 $\pm$ 0.01 & -180.2 $\pm$ 0.3 & -1.7 $\pm$ 1.5 & -1.9 $\pm$ 0.9 & 53.5 & $-$2.87 $\pm$ 0.12 & -2.23 $\pm$ 0.11 &  Y  \\ \\ 
298.15270 & $-$22.08463 & 19.25 $\pm$ 0.01 & 18.46 $\pm$ 0.01 & 20.11 $\pm$ 0.01 & -176.2 $\pm$ 0.5 & 12.3 $\pm$ 0.5 & -27.2 $\pm$ 0.3 & 27.6 & $-$2.83 $\pm$ 0.12 & -2.23 $\pm$ 0.11 &  Y  \\ \\ 
298.15185 & $-$22.09654 & 19.29 $\pm$ 0.01 & 18.52 $\pm$ 0.01 & 20.18 $\pm$ 0.01 & -177.6 $\pm$ 0.4 & 6.8 $\pm$ 3.3 & 0.1 $\pm$ 2.1 & 29.0 & $-$2.56 $\pm$ 0.12 & -2.25 $\pm$ 0.12 &  Y  \\ \\ 
298.18099 & $-$22.20728 & 19.58 $\pm$ 0.01 & 18.99 $\pm$ 0.01 & 20.55 $\pm$ 0.02 & 35.3 $\pm$ 0.8 & --- $\pm$ --- & --- $\pm$ --- & 11.7 & $-$1.14 $\pm$ 0.12 & --- &  N  \\ \\ 
298.21784 & $-$22.06409 & 19.66 $\pm$ 0.01 & 18.90 $\pm$ 0.01 & 20.47 $\pm$ 0.02 & -175.1 $\pm$ 0.4 & -2.4 $\pm$ 0.7 & -5.2 $\pm$ 0.4 & 34.3 & $-$99.0 $\pm$ 0.00 & -2.36 $\pm$ 0.1 &  Y  \\ \\ 
298.16019 & $-$22.09027 & 19.94 $\pm$ 0.01 & 19.34 $\pm$ 0.01 & 20.77 $\pm$ 0.02 & 93.9 $\pm$ 2.8 & -0.3 $\pm$ 0.8 & -1.3 $\pm$ 0.4 & 3.6 & $-$1.70 $\pm$ 0.12 & --- &  N  \\ \\ 
298.19280 & $-$22.05979 & 19.99 $\pm$ 0.01 & 19.28 $\pm$ 0.01 & 20.72 $\pm$ 0.02 & -174.3 $\pm$ 0.6 & -0.8 $\pm$ 0.3 & -1.0 $\pm$ 0.1 & 22.6 & $-$3.07 $\pm$ 0.12 & -2.3 $\pm$ 0.11 &  Y  \\ \\ 
298.19388 & $-$22.07918 & 20.01 $\pm$ 0.01 & 19.28 $\pm$ 0.01 & 20.79 $\pm$ 0.02 & -176.2 $\pm$ 0.6 & -7.1 $\pm$ 1.0 & -8.9 $\pm$ 0.6 & 17.7 & $-$2.67 $\pm$ 0.12 & -2.15 $\pm$ 0.13 &  Y  \\ \\ 
298.20960 & $-$22.07638 & 20.03 $\pm$ 0.01 & 19.33 $\pm$ 0.01 & 20.87 $\pm$ 0.02 & 121.4 $\pm$ 1.3 & 0.0 $\pm$ 0.5 & -0.4 $\pm$ 0.3 & 22.5 & $-$2.26 $\pm$ 0.12 & -1.78 $\pm$ 0.13 &  N  \\ \\ 
298.16641 & $-$22.06769 & 20.21 $\pm$ 0.01 & 19.51 $\pm$ 0.01 & 20.95 $\pm$ 0.02 & -172.2 $\pm$ 1.5 & -3.2 $\pm$ 0.6 & -4.5 $\pm$ 0.3 & 10.8 & $-$2.73 $\pm$ 0.12 & -2.72 $\pm$ 0.1 &  Y  \\ \\ 
298.13014 & $-$22.14273 & 20.24 $\pm$ 0.01 & 19.56 $\pm$ 0.01 & 21.01 $\pm$ 0.02 & -178.5 $\pm$ 1.0 & --- $\pm$ --- & --- $\pm$ --- & 14.5 & $-$2.40 $\pm$ 0.12 & -2.06 $\pm$ 0.14 &  Y  \\ \\ 
298.14121 & $-$22.06475 & 20.28 $\pm$ 0.01 & 19.62 $\pm$ 0.01 & 21.05 $\pm$ 0.02 & -175.2 $\pm$ 1.2 & -14.8 $\pm$ 1.5 & -20.4 $\pm$ 0.8 & 15.1 & $-$99.0 $\pm$ 0.00 & -1.89 $\pm$ 0.16 &  Y  \\ \\ 
298.19401 & $-$22.08638 & 20.38 $\pm$ 0.01 & 19.70 $\pm$ 0.01 & 21.09 $\pm$ 0.03 & -175.7 $\pm$ 1.2 & -1.2 $\pm$ 0.9 & -1.1 $\pm$ 0.5 & 7.7 & $-$2.78 $\pm$ 0.12 & -2.17 $\pm$ 0.11 &  Y  \\ \\ 
298.21019 & $-$22.05475 & 20.51 $\pm$ 0.01 & 19.82 $\pm$ 0.01 & 21.20 $\pm$ 0.03 & -188.7 $\pm$ 5.8 & -1.5 $\pm$ 0.5 & -0.8 $\pm$ 0.3 & 3.8 & $-$3.06 $\pm$ 0.12 & --- &  Y  \\ \\ 
298.14290 & $-$22.07784 & 20.53 $\pm$ 0.01 & 19.87 $\pm$ 0.01 & 21.23 $\pm$ 0.03 & -177.4 $\pm$ 1.3 & -1.4 $\pm$ 1.5 & -9.8 $\pm$ 0.7 & 14.3 & $-$2.68 $\pm$ 0.12 & --- &  Y  \\ \\ 
298.17751 & $-$22.16337 & 20.60 $\pm$ 0.01 & 19.93 $\pm$ 0.01 & 21.36 $\pm$ 0.03 & 89.3 $\pm$ 0.8 & -7.3 $\pm$ 0.4 & -10.1 $\pm$ 0.2 & 13.4 & $-$2.38 $\pm$ 0.12 & --- &  N  \\ \\ 
298.21550 & $-$22.08104 & 20.65 $\pm$ 0.01 & 20.00 $\pm$ 0.01 & 21.36 $\pm$ 0.03 & -177.4 $\pm$ 1.0 & -1.4 $\pm$ 1.0 & -1.2 $\pm$ 0.6 & 12.1 & $-$2.60 $\pm$ 0.12 & --- &  Y  \\ \\ 
298.14837 & $-$22.04551 & 20.64 $\pm$ 0.01 & 19.99 $\pm$ 0.01 & 21.26 $\pm$ 0.03 & -176.7 $\pm$ 1.3 & 1.0 $\pm$ 1.3 & 0.6 $\pm$ 0.7 & 12.6 & $-$99.0 $\pm$ 0.00 & --- &  Y  \\ \\ 
298.13415 & $-$22.04367 & 21.03 $\pm$ 0.01 & 20.46 $\pm$ 0.01 & 21.58 $\pm$ 0.04 & -267.1 $\pm$ 2.5 & -1.0 $\pm$ 1.5 & -2.2 $\pm$ 0.9 & 5.3 & $-$2.99 $\pm$ 0.12 & --- &  N  \\ \\ 
298.11950 & $-$22.0253 & 21.19 $\pm$ 0.01 & 20.56 $\pm$ 0.01 & 21.89 $\pm$ 0.05 & -279.0 $\pm$ 2.8 & -2.0 $\pm$ 2.4 & -7.7 $\pm$ 1.4 & 6.7 & $-$2.44 $\pm$ 0.12 & --- &  N  \\ \\ 
298.14955 & $-$22.10703 & 21.24 $\pm$ 0.01 & 20.62 $\pm$ 0.01 & 21.89 $\pm$ 0.05 & -174.5 $\pm$ 11.9 & 4.8 $\pm$ 3.1 & 2.9 $\pm$ 1.8 & 3.4 & $-$2.69 $\pm$ 0.12 & --- &  Y  \\ \\ 
298.03665 & $-$22.04132 & 18.13 $\pm$ 0.01 & 17.23 $\pm$ 0.01 & 19.54 $\pm$ 0.01 & 127.2 $\pm$ 0.5 & 0.2 $\pm$ 3.5 & -1.4 $\pm$ 1.8 & 105.2 & $-$1.23 $\pm$ 0.12 & -1.19 $\pm$ 0.17 &  N  \\ \\ 
298.09544 & $-$22.14968 & 19.71 $\pm$ 0.01 & 18.89 $\pm$ 0.01 & 20.98 $\pm$ 0.02 & -2.7 $\pm$ 0.9 & 0.1 $\pm$ 1.4 & -1.9 $\pm$ 0.7 & 31.0 & $-$1.25 $\pm$ 0.12 & -1.36 $\pm$ 0.15 &  N  \\ \\ 
298.06312 & $-$22.11093 & 20.13 $\pm$ 0.01 & 19.44 $\pm$ 0.01 & 21.05 $\pm$ 0.02 & -29.5 $\pm$ 1.1 & -0.3 $\pm$ 0.8 & -1.3 $\pm$ 0.4 & 12.8 & $-$99.0 $\pm$ 0.00 & -1.37 $\pm$ 0.17 &  N  \\ \\ 
298.05710 & $-$22.02942 & 20.59 $\pm$ 0.01 & 20.02 $\pm$ 0.01 & 21.35 $\pm$ 0.03 & -63.9 $\pm$ 0.8 & 2.7 $\pm$ 1.8 & 0.4 $\pm$ 1.0 & 10.9 & $-$1.89 $\pm$ 0.12 & --- &  N  \\ \\ 
297.96338 & $-$22.04654 & 20.35 $\pm$ 0.01 & 19.76 $\pm$ 0.01 & 21.11 $\pm$ 0.02 & -18.8 $\pm$ 1.5 & --- $\pm$ --- & --- $\pm$ --- & 12.1 & $-$1.99 $\pm$ 0.12 & -1.61 $\pm$ 0.16 &  N  \\ \\ 
298.15344 & $-$22.04959 & 17.76 $\pm$ 0.01 & 16.74 $\pm$ 0.01 & 18.98 $\pm$ 0.01 & -175.5 $\pm$ 0.3 & -0.9 $\pm$ 0.4 & -1.0 $\pm$ 0.2 & 125.2 & $-$2.31 $\pm$ 0.12 & -2.23 $\pm$ 0.12 &  Y  \\ \\ 
298.28018 & $-$22.12346 & 17.18 $\pm$ 0.01 & 16.40 $\pm$ 0.01 & 18.60 $\pm$ 0.01 & -14.2 $\pm$ 0.4 & -0.3 $\pm$ 1.3 & 0.8 $\pm$ 0.8 & 131.5 & --- & -1.19 $\pm$ 0.18 &  N  \\ \\ 
298.01474 & $-$21.91537 & 17.29 $\pm$ 0.01 & 16.29 $\pm$ 0.01 & 18.90 $\pm$ 0.01 & -48.0 $\pm$ 0.4 & -4.0 $\pm$ 1.7 & -8.8 $\pm$ 1.0 & 131.5 & --- & -1.23 $\pm$ 0.18 &  N  \\ \\

\end{tabular}
\end{sideways}
\end{table*}

\newcommand{\mnras}{MNRAS}
\newcommand{\pasa}{PASA}
\newcommand{\nat}{Nature}
\newcommand{\araa}{ARAA}
\newcommand{\aj}{AJ}
\newcommand{\apj}{ApJ}
\newcommand{\apjl}{ApJ}
\newcommand{\apjs}{ApJSupp}
\newcommand{\aap}{A\&A}
\newcommand{\aaps}{A\&ASupp}
\newcommand{\pasp}{PASP}
\newcommand{\pasj}{PASJ}

%\bibliography{/Users/longeard/Documents/biblio}
%\bibliographystyle{mn2e}

\clearpage

\end{document}